\documentclass[11pt]{article}

\usepackage[a4paper,margin=1in]{geometry}
\usepackage{amsmath,amssymb}
\usepackage{graphicx}
\usepackage{booktabs}
\usepackage{tabularx}
\usepackage{makecell}
\usepackage{array}
\usepackage{multirow}
\usepackage[numbers,sort&compress]{natbib}
\usepackage{xcolor}
\usepackage[colorlinks=true,linkcolor=blue,citecolor=blue,urlcolor=blue]{hyperref}
\usepackage{textcomp}

\newcolumntype{Y}{>{\raggedright\arraybackslash}X}

\newcommand{\micron}{\ensuremath{\mu\mathrm{m}}}
\newcommand{\microsec}{\ensuremath{\mu\mathrm{s}}}
\newcommand{\GBP}{\pounds}

\title{A Backend-Agnostic MWIS Kernel for Stochastic Unit Commitment with Neutral-Atom Hardware Validation}

\author{%
Jiying Chen$^{1,\ast}$, Min Lin$^{2}$, Jingwei Wen$^{3}$, Zhihong Zhang$^{3}$, Chuixiong Wu$^{4,\ast}$\\[4pt]
\small $^{1}$Department of Chemical Engineering and Biotechnology, University of Cambridge, Cambridge, UK\\
\small $^{2}$The Cavendish Laboratory, University of Cambridge, Cambridge, UK\\
\small $^{3}$China Mobile (Suzhou) Software Technology Co., Ltd., Suzhou, China\\
\small $^{4}$Taiyi Quantum Science \& Technology Co. Ltd., Shanghai, China\\
\small $^{\ast}$Corresponding authors. e-mail: \texttt{jc2341@cam.ac.uk} (J. Chen); \texttt{cxwu@taiyiliangsheng.com} (C. Wu)
}
\date{}

\begin{document}

\maketitle

\begin{abstract}
Quantum hardware is beginning to address structured combinatorial optimisation, but two steps still block practical use: mapping real operational models onto hardware-compatible instances, and converting noisy hardware output back into feasible decisions. Here we introduce a backend-agnostic computational interface that compiles the discrete decision layer of stochastic unit commitment into a move-based maximum-weight independent set (MWIS) problem, while retaining continuous dispatch and feasibility recovery in the classical computational layer. We validate the approach in a green hydrogen scheduling setting and deploy it on the QuEra Aquila neutral-atom quantum processor. This is the first end-to-end industrial scheduling workflow that connects real operational decisions to programmable neutral-atom hardware through a solver-agnostic MWIS representation. Across a 15-day hardware campaign on 50-node instances, hardware-generated solutions after classical refinement match or exceed the dispatch margins obtained from exact MWIS on every day. When scaling to 144 nodes, encoding quality remains stable, while the probability that the full atom array survives, rather than graph embedding, emerges as the dominant bottleneck to further scaling. Together, these results establish a hardware-compatible computational pathway toward larger problem scales, and lay the groundwork for exploring regimes in which exact classical optimisation may no longer scale efficiently.
\end{abstract}

\section{Introduction}\label{sec:intro}

Stochastic unit commitment (UC) asks, for each of $N$ controllable units across a $T$-hour horizon, which units to commit and at what operating point so that a delivery contract is met under uncertain supply, at minimum electricity cost. The on/off layer is combinatorial; the dispatch layer is a continuous program. For $N=6$ and $T=24$, the naive binary search space is $2^{N\times T} \approx 2.2 \times 10^{43}$ configurations, most of them infeasible under production and inventory constraints.

We treat a six-module green-hydrogen electrolyser park under stochastic wind and solar supply as the industrial case. When and at what load an electrolyser runs sets both cost and the effective carbon intensity of the delivered gas \citep{glenk2019,giovanniello2024}. Electrolyser UC inherits the thermal-generator on/off structure but adds coupled production, storage, and delivery constraints; at plant scale those include thermal and hydrogen-to-oxygen impurity dynamics that a purely electrical model omits \citep{qiu2023}. The kernel below is defined for stochastic UC; the hydrogen park supplies the reference schedules, duals, and dispatch LP used throughout.

UC is NP-hard, with a classical literature that includes Benders-style decomposition of binary commitment from continuous dispatch \citep{benders1962} and tight mixed-integer formulations of the power-based UC problem \citep{morales2015}. Quantum UC work to date has used a two-stage QUBO-plus-LP split on quantum annealers: \citet{barrass2025} applied it to IEEE RTS-96 on D-Wave hardware, and \citet{christeson2025} formulated the binary master problem as a QUBO inside a Benders loop, evaluating instances from 10 to 1{,}000 units. Related studies benchmark D-Wave hybrid solvers against classical MIP on UC-type MILPs \citep{quinton2025} and chance-constrained stochastic UC \citep{ribes2026}. The nearest Aquila result is MaxCut on an IEEE 9-bus maximum-power-section task, on graphs with up to 12 vertices and no unit-commitment or dispatch-repair layer \citep{bauer2026}. MaxCut scores cut edges between two partitions; MWIS scores a conflict-free vertex subset. To our knowledge no prior work has encoded UC as MWIS on neutral-atom hardware with integrated dispatch repair (Table~\ref{tab:positioning}).

\begin{table}[htbp]
\centering
\caption{Positioning against prior quantum UC and Aquila power-system work. This paper's industrial case is a green-hydrogen electrolyser park; the kernel is a stochastic UC compilation.}
\label{tab:positioning}
\footnotesize
\setlength{\tabcolsep}{4pt}
\begin{tabular*}{\textwidth}{@{\extracolsep{\fill}}lllll@{}}
\toprule
Work & Problem & Platform & Scale & Dispatch repair \\
\midrule
\citet{bauer2026} & MaxCut (IEEE 9-bus) & Aquila AHS & $\le 12$ vertices & none \\
\citet{barrass2025} & UC (IEEE RTS-96) & D-Wave & RTS-96 & QUBO + LP \\
\citet{christeson2025} & UC & D-Wave & 10--1{,}000 units & QUBO (Benders) + LP \\
\citet{quinton2025} & UC-type MILP & D-Wave hybrid & MIP benchmark & hybrid vs.\ MIP \\
This work & stochastic UC (H$_2$ case) & Aquila AHS & 50--144 nodes, 15 days & 100-scenario LP \\
\bottomrule
\end{tabular*}
\end{table}

Maximum Weight Independent Set (MWIS), Quadratic Unconstrained Binary Optimization (QUBO), and Analog Hamiltonian Simulation (AHS) are standard encodings for graph-structured combinatorial problems \citep{lucas2014}. MWIS on unit-disk graphs maps to the ground state of a Rydberg array under blockade; QUBO is the native format of quantum annealers and coherent Ising machines \citep{inagaki2016,hamerly2019,honjo2021}. Using QUBO or MWIS as a shared intermediate representation is established practice \citep{wurtz2024hybrid}. The claim here is a concrete conflict-graph MWIS compilation of stochastic UC---move vocabulary, LP-dual weights, folding embedding---run end-to-end on Aquila, and in principle mappable to annealing or a coherent Ising machine without changing the kernel.

Rydberg blockade encodes the independent-set constraint directly: two atoms within radius $R_b$ cannot occupy the excited state simultaneously. \citet{pichler2018} showed that MIS on unit-disk graphs is the ground state of a Rydberg array with zero ancilla overhead. \citet{ebadi2022} demonstrated unweighted MIS on programmable 2D arrays with up to 289 atoms and reported a superlinear scaling advantage over simulated annealing on high-degeneracy graphs. Weighted MWIS has since been encoded with local light shifts, first in simulation \citep{goswami2024} and then experimentally \citep{deoliveira2025}; vertex-support-derived local detunings have likewise been used to bias MIS preparation toward better approximations \citep{yeo2025}. Improved classical simulated-annealing protocols close much of that speedup gap \citep{andrist2023}; we therefore use simulated annealing as a calibration reference, not as a solver to beat. King-lattice MIS datasets now exist at the hundred-atom scale \citep{kim2024data}. Compilation toolkits that reduce generic MIS instances onto Rydberg-native unit-disk graphs report an easy--hard--easy transition in native instance hardness \citep{schuetz2025}.

\citet{nguyen2023} encode arbitrary graphs with copy, crossing, and crossing-with-edge gadgets on a square crossing lattice. The construction is universal but costs $O(N^2)$ ancillae for $N$ logical variables, which is prohibitive unless the graph is already unit-disk. Sparse non-local edges can use Rydberg quantum wires \citep{kim2022wires,deoliveira2026wires}; dense or higher-order constraints can use a parity-to-MWIS architecture \citep{lanthaler2023}. Gadget mappings can close the spectral gap even when the original instance is easy \citep{bombieri2025}. Aquila's hardware envelope---256 atoms, FOV $75\times76$~\micron, $R_b = 8$--$10$~\micron---constrains any practical embedding; static traps are filled with $\sim$60\% probability, so usable shots are post-selected on fully loaded arrays \citep{wurtz2023}.

\citet{pan2025} showed that the $V \propto 1/r^6$ tail degrades square-lattice King's subgraph (KSG) encodings and proposed triangular-lattice subgraph (TLSG) encoding as a higher-fidelity alternative, at the cost of more complex gadgets. For temporally extended UC, folding one lattice axis at the split that minimises occupied span trades a bounded set of broken inter-row edges for FOV compliance, without gadgets.

On Aquila, we place each AHS-eligible move at a distinct (module, hour) cell with Chebyshev-distance-$\le 1$ conflicts---an 8-neighbour King grid that is natively unit-disk (Section~\ref{sec:embedding}). A 24-hour horizon exceeds the FOV (15 hours at $s = 5$~\micron), so a flat layout fails. Two-row folding splits the time axis so that each row's occupied hour-span is minimised, and greedy post-processing repairs broken inter-row edges. The full 24-hour instance then embeds in a single AHS shot with zero gadget overhead.

The kernel is a \textbf{move-MWIS compilation} of stochastic UC. From one feasible reference commitment we build a vocabulary of locally valid schedule perturbations (moves). An MWIS solver---exact or approximate---selects a compatible subset; classical LP repair then validates the modified commitment. The construction is close in spirit to local branching around a MIP incumbent \citep{fischetti2003}, except that the neighbourhood is compiled into a weighted conflict graph rather than searched inside a branch-and-cut solver. The hydrogen electrolyser park is the case on which the kernel is executed end-to-end. AHS is the first hardware backend; quantum annealing and coherent Ising machines can use the same graph.

This paper reports three results. First, the pipeline runs UC$\to$MWIS$\to$Aquila$\to$100-scenario LP repair as a closed loop, which prior Aquila power-system work does not. Second, a 15-day cloud campaign at $N=50$ yields hybrid AHS+SA dispatch margins that match or exceed exact MWIS+LP on every day. Third, as $N$ grows the valid-shot rate, not FOV occupancy, becomes the binding AHS scale limit. The supporting computational pieces, written as a transferable kernel, are LP-dual move scoring from a single stochastic LP, 2-row folding with zero ancillae, and simulated-annealing refinement on the true dispatch margin rather than the MWIS proxy.

\section{Methods --- Architecture}\label{sec:methods}

\subsection{The Hybrid Combinatorial Pipeline}\label{sec:pipeline}

The kernel compiles a single-day stochastic UC reference into a weighted conflict graph and a dispatch-repair LP. Experiments use a six-module, 24-hour green-hydrogen instance; the stages are instance-agnostic. The pipeline has six core stages, plus a downstream SA local-refinement pass applied to the graph-solver output before dispatch:

\paragraph{Reference schedule.} A baseline feasible commitment is constructed by chronological fill, cheapest fill, or a pre-existing artefact. This is the starting point that the excitation active space perturbs.

\begin{figure}[htbp]
\centering
\includegraphics[width=\linewidth]{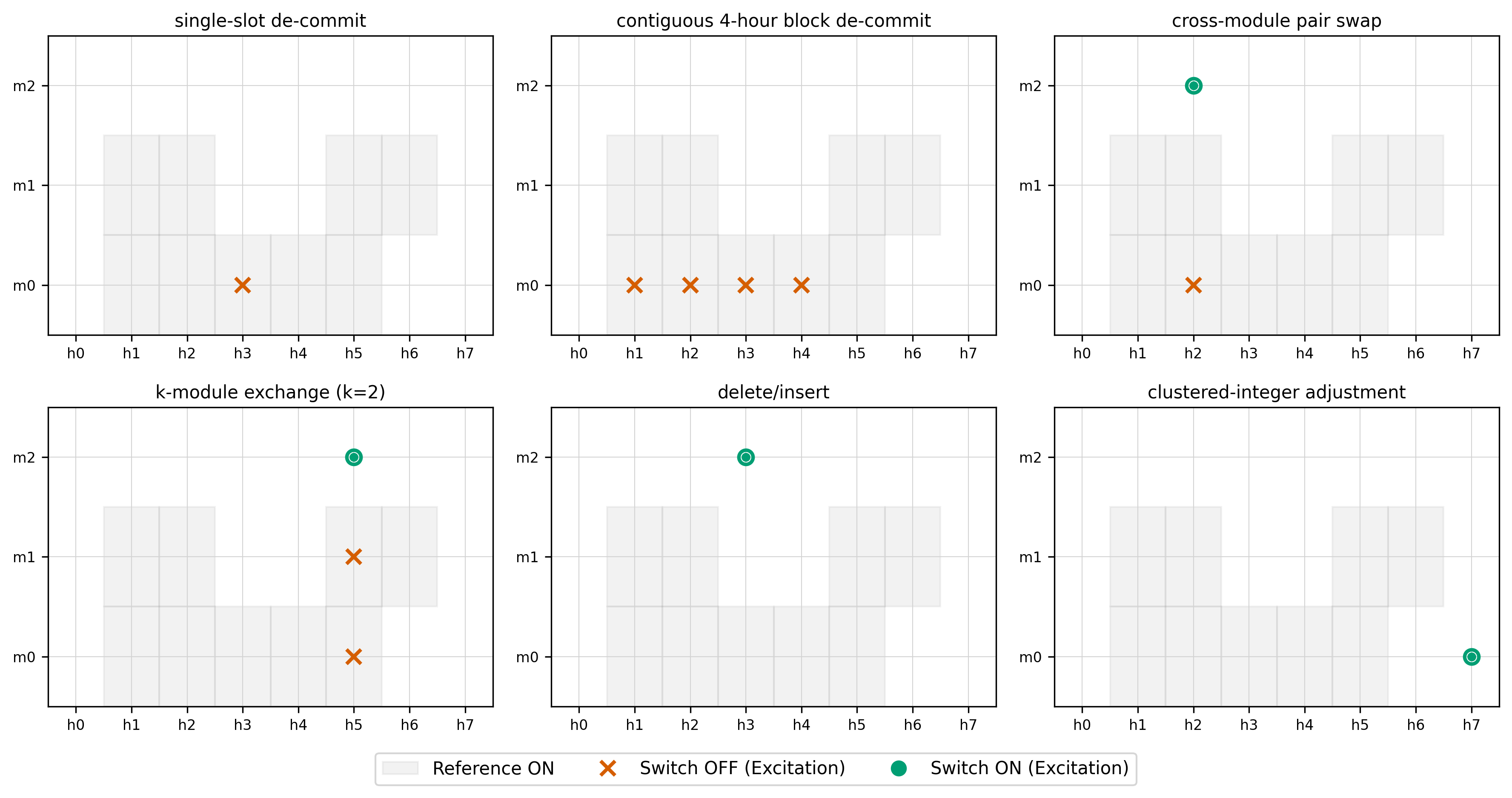}
\caption{Move vocabulary---six move types populating the excitation active space, covering single-slot, block, cross-module, exchange, delete/insert, and clustered-integer perturbations.}
\label{fig:moves}
\end{figure}

\paragraph{Move generation.} Six move types populate the excitation active space: single-slot de-commit, contiguous 4-hour block de-commit, cross-module pair swap, $k$-module exchange, delete/insert, and clustered-integer neighbourhood adjustment (Figure~\ref{fig:moves}; Supplementary Section~S5). Each move is a well-defined local perturbation of the reference schedule: a set of slots to switch off and a set of slots to switch on.

\paragraph{LP-dual scoring.} Each move is scored using the dual values (shadow prices) from a single classical LP solve on the reference schedule. The LP captures coupling between electricity cost, tank inventory, grid load, delivery contracts, and terminal boundary conditions. Writing $\lambda_t^{C5}$ and $\lambda_t^{C7}$ for the probability-weighted, sign-flipped marginals of the power-balance and hydrogen-tank-dynamics equality constraints at hour $t$,
\begin{equation}
\lambda_t^{C5} = -\sum_{s=1}^{S} \pi_s \, \mu_{t,s}^{C5}, \qquad \lambda_t^{C7} = -\sum_{s=1}^{S} \pi_s \, \mu_{t,s}^{C7}
\label{eq:duals}
\end{equation}
where $\pi_s$ are scenario probabilities and $\mu_{t,s}^{C5}, \mu_{t,s}^{C7}$ are the raw minimisation-sense LP marginals, a move $m$'s score is the linear combination
\begin{equation}
\mathrm{score}(m) = \sum_{t} \Big[ P_{\mathrm{EL}} \, \lambda_t^{C5} + \eta_{\mathrm{EL}} P_{\mathrm{EL}} \, \lambda_t^{C7} \Big] \, \Delta_t^{\mathrm{on}}(m)
\label{eq:score}
\end{equation}
with $P_{\mathrm{EL}} = 5.0$~MW (per-module electrolyser power), $\eta_{\mathrm{EL}} = 18.5$~kg/MWh (electrolyser hydrogen yield), and $\Delta_t^{\mathrm{on}}(m) \in \{-1, 0, +1\}$ the move's ON-indicator delta at hour $t$ (e.g.\ for a pair-swap moving capacity from hour $t_{\mathrm{off}}$ to hour $t_{\mathrm{on}}$, the score reduces to the difference form $P_{\mathrm{EL}}(\lambda_{t_{\mathrm{on}}}^{C5} - \lambda_{t_{\mathrm{off}}}^{C5}) + \eta_{\mathrm{EL}} P_{\mathrm{EL}}(\lambda_{t_{\mathrm{on}}}^{C7} - \lambda_{t_{\mathrm{off}}}^{C7})$). This dual score provides a first-order estimate of how the move affects the LP objective, substantially outperforming the naive electricity-price proxy that failed in earlier experiments (35\% sign agreement with ground-truth margin direction).

Scoring is cheap by construction: one continuous LP solve---the reference schedule's 100-scenario stochastic dispatch LP, commitments held fixed and solved once with HiGHS---yields the dual arrays $\lambda^{C5}, \lambda^{C7}$. Every candidate is then scored by Eq.~\eqref{eq:score}, a handful of floating-point operations per move. No per-move LP re-optimisation occurs at scoring time. The scoring stage thus reduces to a single LP solve; the combinatorial UC search is never enumerated. A separate development check---re-solving the full LP once per individual move---was used to validate the dual-score approximation against the true per-move margin delta; that check is not part of the reported pipeline.

\paragraph{Diversity top-$k$ pruning.} Moves are ranked by their absolute LP-dual score, and the top $k = 50$ are retained (Section~\ref{sec:nscaling} revisits this cutoff directly as the $N$-scaling parameter). A diversity filter prevents concentration on the same module-hour slot: at most a bounded number of moves may target any given slot, ensuring the MWIS solver receives a spread of candidate perturbations rather than redundant variants.

\paragraph{Exact MWIS via HiGHS MILP.} The pruned move set is encoded as a conflict graph $G=(V,E)$, where $V$ is the set of retained moves and $(u,v) \in E$ iff moves $u$ and $v$ touch a shared (module, hour) slot. Node weights $w_v = \mathrm{score}(v)$ are the LP-dual scores from Eq.~\eqref{eq:score}. The Maximum Weight Independent Set problem is
\begin{equation}
\max_{x \in \{0,1\}^{|V|}} \ \sum_{v \in V} w_v \, x_v \quad \text{s.t.} \quad x_u + x_v \le 1 \ \ \forall (u,v) \in E
\label{eq:mwis}
\end{equation}
solved to optimality using the HiGHS mixed-integer linear programming solver \citep{huangfu2018}, selecting a compatible set of moves that maximises the total LP-dual proxy objective. On hardware (Section~\ref{sec:embedding}), the same move nodes are placed on a King-graph unit-disk layout and submitted to AHS, which approximates the MWIS objective through Rydberg blockade rather than exact branch-and-bound.

\paragraph{SA local refinement.} The MWIS-selected move set---from HiGHS in the classical pipeline, or from the post-processed AHS selection in the hardware pipeline (Section~\ref{sec:embedding})---is passed as the \emph{warm-start seed} to a simulated annealing solver that searches the same conflict-graph neighbourhood for a higher-value compatible set. SA refines the graph-solver output; it is not a substitute for HiGHS or AHS. As a control, a \emph{cold-start} SA run---seeded from an empty move set---is evaluated alongside the classical HiGHS path (Section~\ref{sec:classical}); the gap between cold-start and warm-start measures how sharp the LP-dual score landscape is.

\begin{figure}[htbp]
\centering
\includegraphics[width=\linewidth]{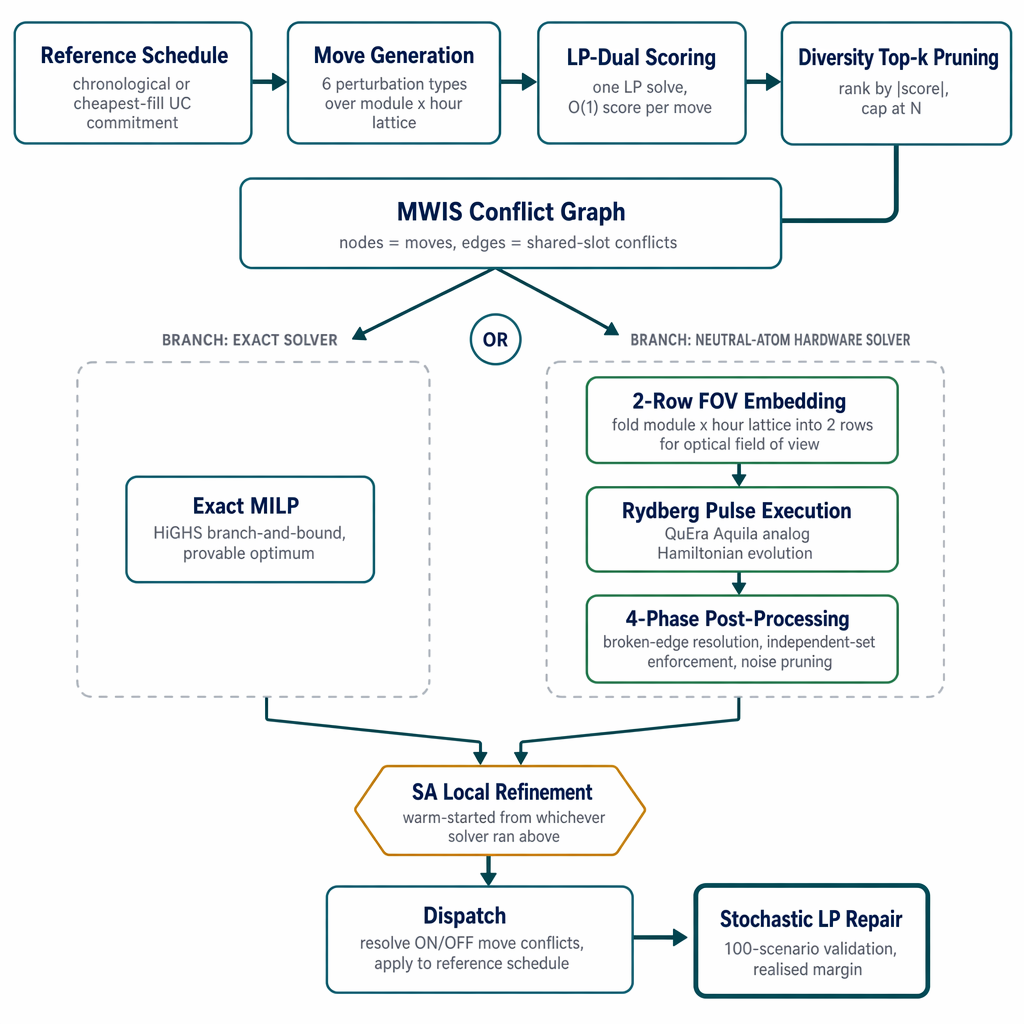}
\caption{Hybrid pipeline architecture---reference schedule, move generation, LP-dual scoring, diversity top-$k$ pruning, an exact-solver/neutral-atom-hardware branch point (exact MILP via HiGHS, or AHS hardware with 2-row FOV embedding and 4-phase post-processing, Section~\ref{sec:embedding}), converging into a common SA local refinement stage, dispatch application, and 100-scenario stochastic LP repair validation.}
\label{fig:pipeline}
\end{figure}

\paragraph{Dispatch and LP repair.} The SA-refined moves are applied to the reference schedule by a dispatch solver: for each module-hour slot, if a selected OFF move and a selected ON move both target the slot, dispatch resolves the contradiction. The resulting modified schedule is validated through a 100-scenario stochastic LP repair that verifies delivery contract feasibility and computes the realised LP margin. The LP repair is a read-only gold standard; its output is never consulted during move generation or MWIS selection. Figure~\ref{fig:pipeline} summarises the architecture.

\subsection{Conflict-graph intermediate representation}\label{sec:ir}

The kernel emits a weighted conflict graph: nodes are moves, weights are LP duals, and edges encode incompatibility. That graph is the intermediate representation demonstrated end-to-end on Aquila. The same instance, with the kernel unchanged, can be handed to several solver classes:

\begin{itemize}
\item \textbf{AHS (neutral-atom):} The conflict graph can be embedded into a unit-disk geometry (with or without gadget overhead) and submitted to a neutral-atom quantum processor that natively solves MWIS via Rydberg blockade.
\item \textbf{QA (quantum annealing):} The same graph can be converted to a QUBO matrix and submitted to a D-Wave quantum annealer, where the quadratic penalty terms enforce the independent-set constraint \citep{barrass2025,quinton2025}.
\item \textbf{CIM (coherent Ising machine):} The same independent-set instance can be mapped to an Ising Hamiltonian and submitted to a CIM, which finds low-energy spin configurations through degenerate optical parametric oscillation \citep{inagaki2016,honjo2021}. CIMs have already been used to compute independent sets on dense graphs with up to 40{,}000 nodes \citep{takesue2025}.
\end{itemize}

The UC encoding is independent of the solver. Move generation, LP-dual scoring, and diversity pruning are classical. Only the graph-solving step---exact MILP today, AHS as the validated alternative, quantum annealing or coherent Ising machines as future targets---depends on the backend; SA refinement (Section~\ref{sec:pipeline}) is common to every backend.

\subsection{Hardware-Adapted Embedding: 2-Row Folding}\label{sec:embedding}

The classical pipeline (Section~\ref{sec:pipeline}) scores and prunes the full six-type move vocabulary, with conflicts defined by slot-sharing: two moves conflict iff they touch an overlapping (module, hour) cell. The AHS pathway narrows this in two ways. First, only single-cell moves---one atom per (module, hour) pair, with an off (de-commit) or on (insert) direction---are AHS-eligible; the multi-cell types (contiguous block de-commit, cross-module pair swap, $k$-module exchange) cannot be placed at a single lattice site and remain in the classical vocabulary (Section~\ref{sec:classical}). Second, because single-cell moves already occupy distinct cells, the slot-sharing rule would leave them unconflicted; the AHS conflict graph instead uses the Chebyshev-distance-$\le 1$ King-graph rule below---a geometric adjacency chosen for native unit-disk embeddability, not because it reproduces a specific UC scheduling constraint. King-lattice MIS is a standard experimental target for Rydberg arrays \citep{ebadi2022,kim2024data}. Only the same-module, adjacent-hour edges have a plausible physical reading, approximating a minimum-downtime constraint between consecutive commitment changes; the remaining edges are a hardware-compatibility proxy. Feasibility of the resulting commitment is not claimed from the King graph. Every dispatch is re-validated against the full 100-scenario stochastic LP---including grid-capacity and tank-dynamics constraints the geometric graph does not encode---before any result is reported (Section~\ref{sec:pipeline}). The selected moves remain a small, local perturbation of an already-feasible reference (the rolling campaign selects 6--13 moves out of the pruned 50, from 144 module-hour cells).

The naive flat 8-neighbour grid embedding of the module$\times$hour lattice is blocked by Aquila's FOV constraint: at $s = 5$~\micron\ atom spacing, the 24-hour horizon requires 120~\micron\ in the $x$-dimension, exceeding the 75~\micron\ hardware limit. The \textbf{2-row folding strategy}, implemented in our published code, compresses the temporal axis by splitting hours into two stacked rows at the point that minimises the larger of the two per-row occupied hour-spans, searched exhaustively over all candidate split points rather than a fixed split by hour count.

Writing $\tau$ for a candidate split hour and $\mathrm{span}(\cdot)$ for the occupied-hour range of a set of (module, hour) cells, the split point is chosen by
\begin{equation}
\tau^{*} = \arg\min_{\tau} \ \max\Big(\mathrm{span}\big(\{(m,h): h < \tau\}\big),\ \mathrm{span}\big(\{(m,h): h \ge \tau\}\big)\Big)
\label{eq:split}
\end{equation}
Each move is then placed at
\begin{equation}
y(m, h) = \begin{cases} m \cdot s, & h < \tau^{*} \quad \text{(row 0)} \\ m \cdot s + (M+1)\cdot s + b, & h \ge \tau^{*} \quad \text{(row 1)} \end{cases}
\qquad
x(m,h) = \big(h - h_{\min,\mathrm{row}}\big) \cdot s
\label{eq:placement}
\end{equation}
where $M$ is the maximum module index, $b = 9$~\micron\ is the additional inter-row separation beyond the regular $5$~\micron\ lattice spacing, and $h_{\min,\mathrm{row}}$ is that row's \textbf{own} minimum occupied hour---not the global minimum across both rows. This per-row $x$-normalisation provides the folding's $x$-axis compression: each row reuses the same $x$ range instead of sharing one continuous axis.

\begin{figure}[htbp]
\centering
\includegraphics[width=\linewidth]{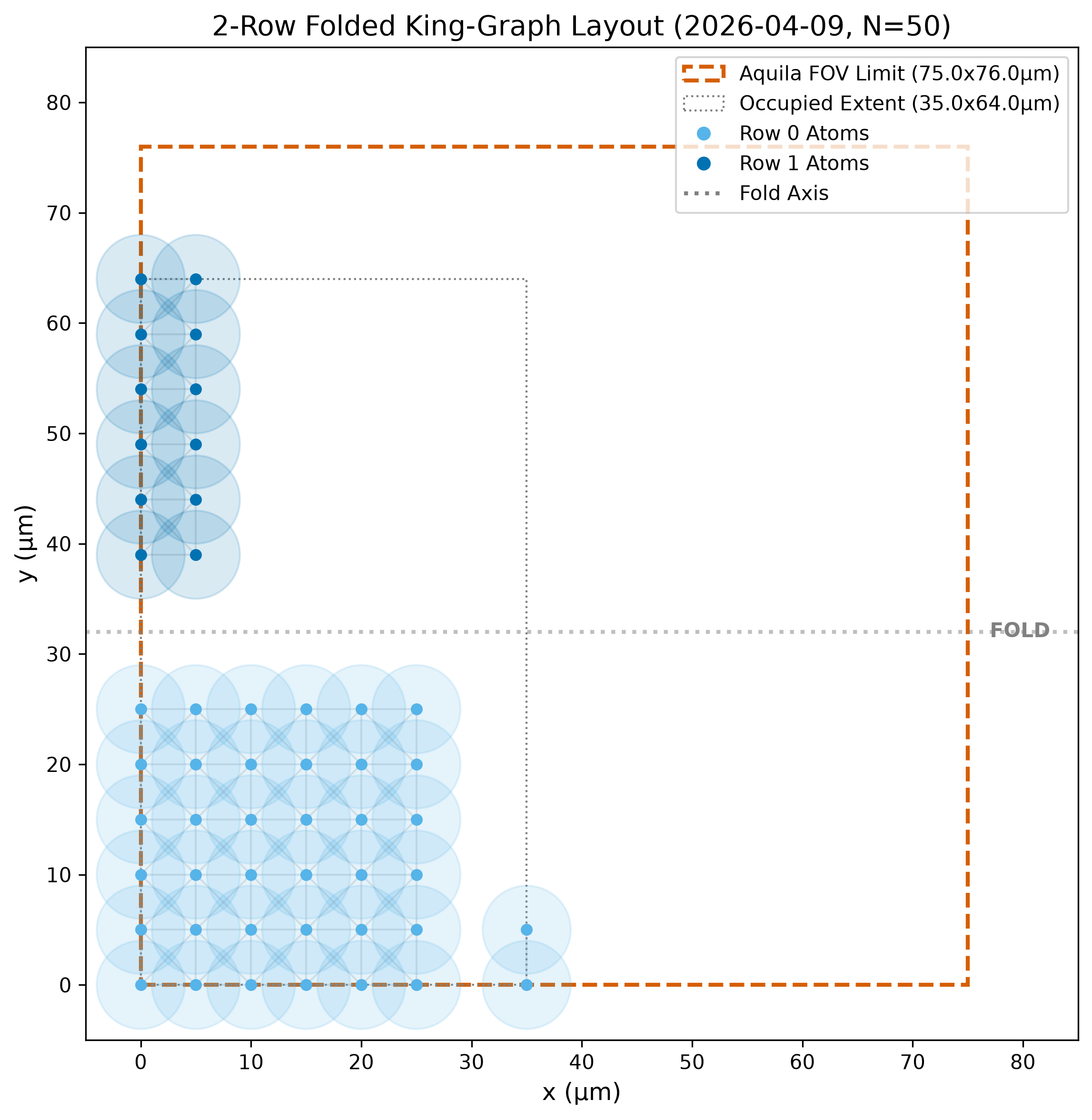}
\caption{2-row folded King-graph layout for the $N=50$ instance on Day 1. The temporal axis is compressed into two stacked rows. Atoms are coloured by row assignment (sky blue for Row 0, blue for Row 1). The orange dashed rectangle marks Aquila's $75\times76$~\micron\ hardware FOV limit; the grey dotted rectangle marks this instance's occupied extent ($35\times64$~\micron)---the layout uses well under half of the available FOV. This day's span-minimising split lands on a natural gap in the occupied-hour distribution, so no inter-row edges are broken. The two atoms isolated at $x\approx35$~\micron\ (bottom right) are an artefact of top-$k$ diversity pruning (Section~\ref{sec:pipeline}): hour 11 survives pruning for two modules while hour 10 does not, leaving a gap in Row 0's occupied-hour coverage.}
\label{fig:folding}
\end{figure}

For a 6-module, 24-hour instance with every cell populated, the split falls at hour 12 (a fully contiguous 24-hour range has no gap to exploit, so the span-minimising search recovers the same split an even count-based split would choose), and this compresses the $x$-dimension from 120~\micron\ to 55~\micron, fitting within the 75~\micron\ FOV. The $y$-dimension reaches 64~\micron\ ($(2M+1)\cdot s + b = 11 \times 5~\micron\ + 9~\micron\ = 64~\micron$), also within the 76~\micron\ limit. The folding is a purely engineering response to the FOV constraint: it preserves the Chebyshev-distance-1 adjacency rule within each row but severs edges between nodes whose (module, hour) cells are within Chebyshev distance $\le 1$ yet lie in different rows. These \textbf{broken edges} violate the independent-set constraint after MWIS selection. Quantum wires offer an alternative encoding of a small number of non-local edges without folding \citep{kim2022wires,deoliveira2026wires}; we instead keep a native King layout and repair the few broken edges classically. Because the split point is chosen to minimise each row's span (Eq.~\eqref{eq:split}), it typically lands on the largest natural gap in the occupied-hour distribution when one exists, which also tends to avoid breaking Chebyshev-adjacent (consecutive-hour) pairs. Figure~\ref{fig:folding} shows the folded $N=50$ layout submitted on Day 1. The Aquila campaign comprises 15 consecutive calendar days (Day 1 to Day 15), with Day 1 corresponding to 9 April 2026. Subsequent text, tables, and figure axes use this Day numbering only.

\paragraph{Post-processing broken edges.} After MWIS selection on the folded graph, our post-processing implementation applies Ebadi-style greedy vertex reduction \citep{ebadi2022}: while any broken-edge pair has both endpoints selected, remove the lower-weight endpoint. This is a lossy step---the resulting set is a subset of the MWIS optimum---but preserves unit-disk feasibility on the folded hardware layout. Across the 15-day rolling campaign (Section~\ref{sec:innerkernel}), the post-processed AHS selection retains a mean 94.0\% of the exact MWIS objective (range 74.8--100\%).

\subsubsection{AHS Schedule Parameter Design}\label{sec:ahsparams}

Before deploying the folded graph on Aquila, the AHS schedule parameters must be tuned to achieve usable MWIS encoding quality. An exhaustive 11-strategy sweep was conducted at $N=12$ (a 2-row folded subgraph with 26 edges, weights $\in [-52.1, 101.9]$) using exact dense diagonalisation ($2^{12} = 4096$ states) at freeze time $t = 0.9T$.

Only 3 of 11 strategies pass the encoding gate, and all three share identical critical parameters (Supplementary Section~S2, Table~S2 and Figure~S1). The sole viable recipe, termed \emph{narrow-zero-cross}, is defined in Table~\ref{tab:recipe}.

\begin{table}[htbp]
\centering
\caption{The \emph{narrow-zero-cross} AHS schedule recipe.}
\label{tab:recipe}
\small
\setlength{\tabcolsep}{5pt}
\begin{tabularx}{\textwidth}{@{}>{\raggedright\arraybackslash}p{0.27\textwidth} >{\raggedright\arraybackslash}p{0.33\textwidth} Y@{}}
\toprule
Parameter & Value & Role \\
\midrule
$\omega_{\max}$ (Rabi frequency)
  & $8 \times 10^{6}$~rad/s (8~MHz)
  & Quantum superposition mixing \\
$\Delta_{\mathrm{start}}$ (global detuning, start)
  & $-3 \times 10^{7}$~rad/s ($-30$~MHz)
  & Temporal encoding sweep: negative phase \\
$\Delta_{\mathrm{end}}$ (global detuning, end)
  & $+3 \times 10^{7}$~rad/s ($+30$~MHz)
  & Temporal encoding sweep: positive phase \\
$\Delta_{\mathrm{local\_peak}}$ (local detuning)
  & $-1.25 \times 10^{8}$~rad/s ($-125$~MHz, hardware maximum)
  & Per-node weight differentiation \\
$T$ (evolution time)
  & $4.0 \times 10^{-6}$~s (4~\microsec)
  & Schedule duration \\
Shape
  & Plateau
  & Sustained peak bias for most of $T$ \\
\bottomrule
\end{tabularx}
\end{table}

$\omega_{\max}$ was held fixed at 8~MHz throughout this 11-strategy sweep---only the global detuning range, local detuning peak, shape, and $T$ were varied (Supplementary Section~S2, Table~S2). A narrow $\pm 1$~MHz sensitivity check around this fixed value is reported separately in Section~\ref{sec:paramtuning} and found no improvement over the baseline.

The evolution is governed by the standard Rydberg AHS Hamiltonian
\begin{equation}
H(t) = \frac{\Omega(t)}{2}\sum_k \sigma_k^x \; - \; \sum_k \delta_k(t)\, n_k \; + \; \sum_{i<j} \frac{C_6}{r_{ij}^6}\, n_i n_j
\label{eq:hamiltonian}
\end{equation}
where $n_k = |r_k\rangle\langle r_k|$ is the Rydberg-state occupation, $\sigma_k^x = |g_k\rangle\langle r_k| + |r_k\rangle\langle g_k|$, and the per-atom effective detuning combines a global sweep and a per-site local term,
\begin{equation}
\delta_k(t) = \Delta_{\mathrm{global}}(t) + h_k \, \Delta_{\mathrm{local}}(t), \qquad h_k = \frac{w_{\max} - w_k}{w_{\max} - w_{\min}} \in [0,1]
\label{eq:detuning}
\end{equation}
with $w_k$ the node's LP-dual weight from Eq.~\eqref{eq:score}, and $C_6$ the Rydberg van-der-Waals coefficient---implicit in the hardware's atom positions, not a tunable schedule parameter. Per-site local detuning is the same physical mechanism used to encode vertex weights in recent weighted Rydberg MWIS work \citep{goswami2024,deoliveira2025,yeo2025}. For \emph{narrow-zero-cross}, $\Omega(t)$ ramps linearly from 0 to $\Omega_{\max}$ over $[0, 0.1T]$, holds the plateau to $0.9T$, then ramps back to 0; $\Delta_{\mathrm{global}}(t)$ sweeps linearly from $\Delta_{\mathrm{start}}$ to $\Delta_{\mathrm{end}}$ across the full duration $T$; and $\Delta_{\mathrm{local}}(t)$ ramps from 0 to $\Delta_{\mathrm{local\_peak}}$ over $[0, 0.1T]$, holds to $0.7T$, then ramps back to 0 by $T$.

\begin{figure}[htbp]
\centering
\includegraphics[width=\linewidth]{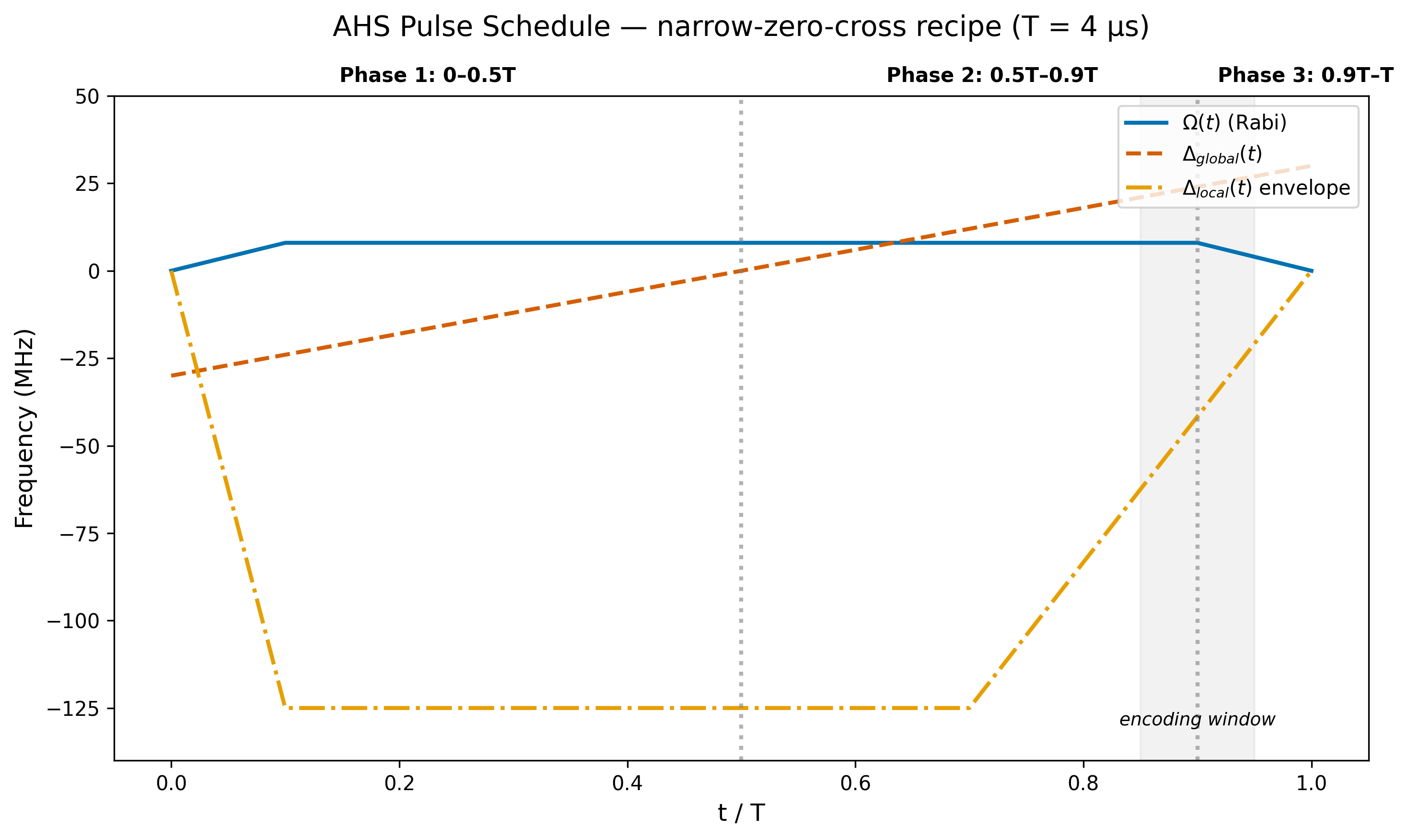}
\caption{AHS pulse schedule for the winning narrow-zero-cross recipe ($T = 4$~\microsec)---Rabi frequency $\Omega(t)$, global detuning $\Delta_{\mathrm{global}}(t)$, and local detuning envelope $\Delta_{\mathrm{local}}(t)$ over normalised time $t/T$, annotated with the three physical phases and the narrow encoding window.}
\label{fig:pulse}
\end{figure}

\begin{figure}[htbp]
\centering
\includegraphics[width=\linewidth]{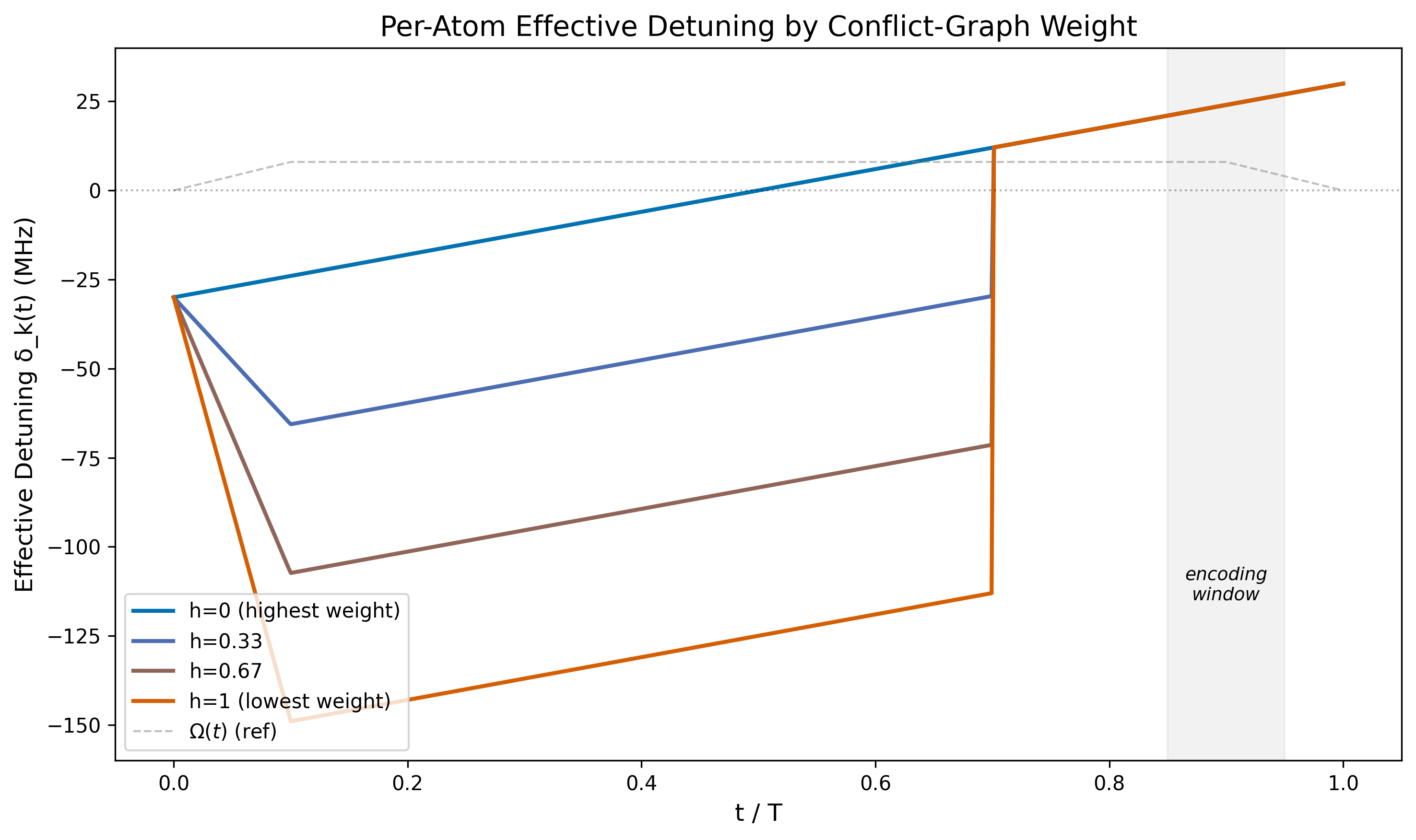}
\caption{Per-atom effective detuning $\delta_k(t) = \Delta_{\mathrm{global}}(t) + h_k\,\Delta_{\mathrm{local}}(t)$ for four representative weight classes ($h_k = 0, 0.33, 0.67, 1$---highest to lowest conflict-graph weight), illustrating the local-detuning bias mechanism of Eq.~\eqref{eq:detuning} under the narrow-zero-cross schedule. High-weight atoms ($h_k\to0$) track $\Delta_{\mathrm{global}}(t)$ alone and cross into positive (excitation-favouring) detuning early in Phase 2; low-weight atoms ($h_k\to1$) retain a strong negative offset that delays their crossing, creating the weight-dependent separation that the freeze at $t/T=0.9$ reads out.}
\label{fig:detuning}
\end{figure}

The physical mechanism operates in three phases. \textbf{Phase 1} ($t = 0$ to $0.5T$): global detuning sweeps from $-30$ to 0~MHz while $\Omega$ ramps to 8~MHz, favouring $|g\rangle$ for all atoms while local detuning ($-125$~MHz peak) differentiates nodes by weight. \textbf{Phase 2} ($t = 0.5T$ to $0.9T$): global detuning crosses zero and becomes positive (reaching $+24$~MHz at freeze), gradually favouring $|r\rangle$; the relative bias from local detuning determines which nodes cross the Rydberg excitation threshold first, with low-weight nodes (more negative local detuning) suppressed longer; high-weight nodes($h_k\rightarrow 0$) cross first. \textbf{Phase 3} ($t = 0.9T$ to $T$): $\Omega$ ramps to zero, freezing the system into the classical Ising Hamiltonian whose low-energy eigenstates approximate MWIS solutions. Figure~\ref{fig:pulse} shows the pulse schedule and Figure~\ref{fig:detuning} the resulting per-atom effective detuning.

The weight-encoding mechanism is visible in the four detuning curves (Figure~\ref{fig:detuning}). During $[0, 0.1T]$, $\Delta_{\mathrm{local}}(t)$ ramps to its full $-125$~MHz peak, fanning the curves apart almost immediately: at $t/T=0.1$, $\delta_k \approx -24 - 125h_k$~MHz, already separated by up to 125~MHz depending on weight. During $[0.1T, 0.7T]$---60\% of the whole schedule---$\Delta_{\mathrm{local}}(t)$ \emph{holds constant} at $-125$~MHz while $\Delta_{\mathrm{global}}(t)$ keeps sweeping upward; every curve therefore moves with the identical slope, maintaining a fixed $125h_k$~MHz separation while all four are pushed uniformly toward zero. This is the ``asking'' phase: since $\Delta_{\mathrm{global}}$ only reaches $+30$~MHz at $t=T$, the highest-weight atom ($h_k=0$, no penalty) is the only one that can cross into positive (excitation-favouring) detuning during this stretch, at $t/T=0.5$ where $\Delta_{\mathrm{global}}$ crosses zero; the lowest-weight atom ($h_k=1$) would need $\Delta_{\mathrm{global}}=+125$~MHz to cross on its own---never reached---so it stays pinned negative throughout the plateau. From $0.7T$ to $T$, $\Delta_{\mathrm{local}}(t)$ retracts linearly to 0, releasing the suppression: all four curves converge and meet exactly at $\delta_k(T) = \Delta_{\mathrm{global}}(T) = +30$~MHz for every $h_k$, since the weight-dependent term vanishes identically once $\Delta_{\mathrm{local}}=0$. This retraction is the ``erasing'' phase---the weight information encoded during the plateau is being actively destroyed as $t\to T$---which is precisely why the freeze must happen partway through this retraction rather than at its end: at $t/T=0.9$, roughly a third of the way through the erasure, the highest- and lowest-weight curves are still separated by about 42~MHz (freeze-point values computed from Eq.~\eqref{eq:detuning}), enough to distinguish them, but that margin is shrinking fast and reaches zero by $t/T=1$.

The eight failing strategies identify necessary conditions: weak local detuning ($-80$~MHz, the old pipeline default) fails across all global sweep configurations (ranks 49--81); an all-negative global sweep ($-90 \to -70$~MHz) suppresses all atoms indiscriminately (rank 68, worst of the plateau family); zero global sweep fails even with hardware-maximum local detuning (ranks 31--35); the triangle shape (no plateau) compounds failure (rank 81 vs.\ rank 57 with plateau at the same detuning values). \textbf{Evolution time $T$ is invariant} within this recipe: ranks and energy gaps are identical at $T = 2$, 4, and 6~\microsec, because the encoding depends on the ratio $t/T$ rather than absolute time, and the adiabatic preparation is sufficient even at the shortest tested duration. The invariance is reported for the \emph{narrow-zero-cross} plateau schedule and its $\sim$200~ns freeze window, not as a general Rydberg result. Related work has used Bayesian optimisation to design annealing schedules on a neutral-atom processor \citep{finzgar2024}; here we instead exhaustively sweep a small, physically motivated schedule family, which is enough to isolate a single viable recipe at $N=12$.

\begin{figure}[htbp]
\centering
\includegraphics[width=\linewidth]{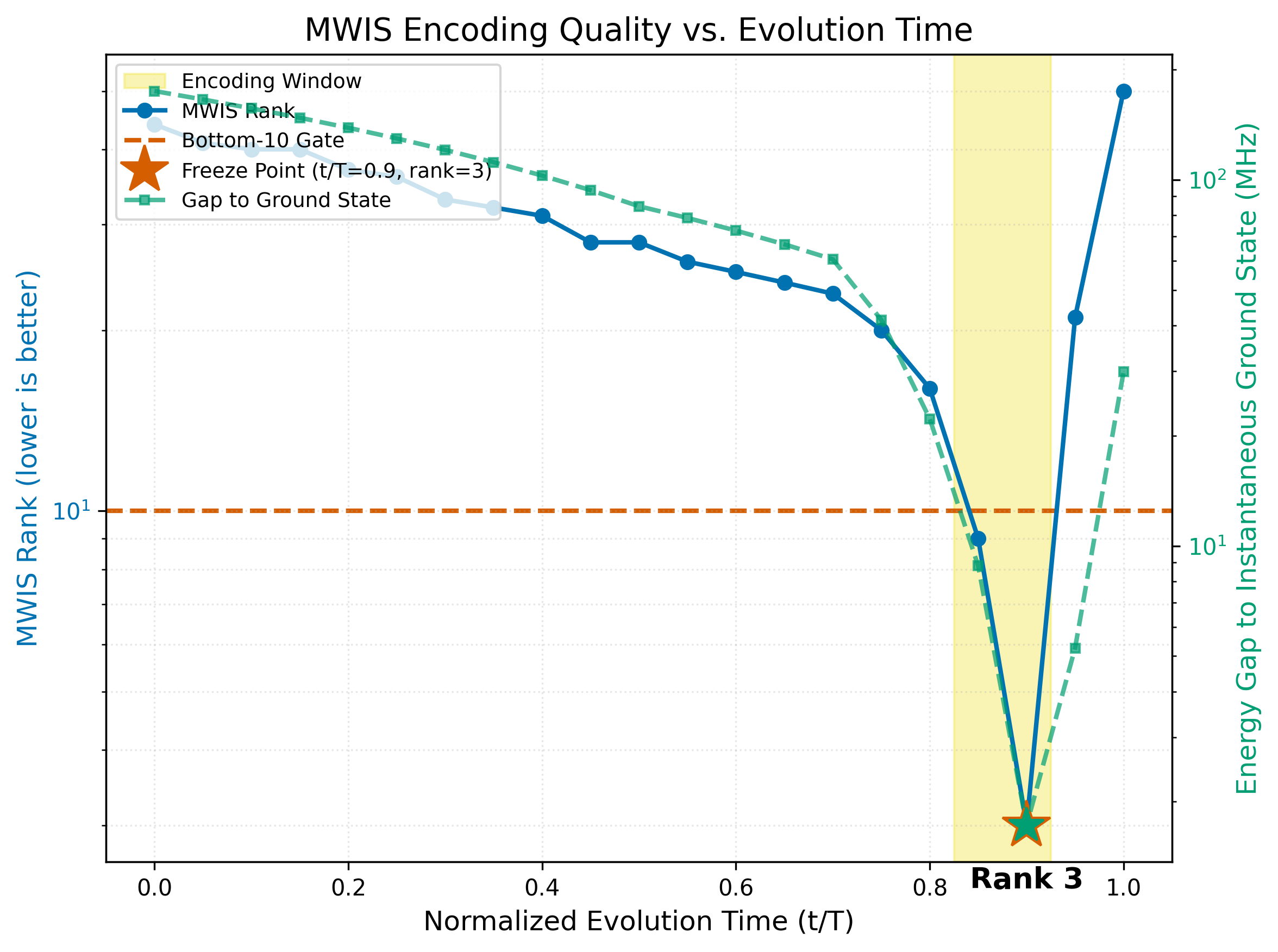}
\caption{MWIS rank (blue, left axis) and the energy gap to the instantaneous ground state (green, right axis, MHz) vs.\ normalised evolution time $t/T$ (exact diagonalisation, $N=12$, no hardware shots). The narrow $\sim$200~ns encoding window near the freeze point is where the target state reaches its best-achieved rank of 3---never rank 1---among the 4096 basis states. The two curves share the same minimum, so the low rank corresponds to a small energy gap rather than a favourable position in a sparse spectrum.}
\label{fig:window}
\end{figure}

\paragraph{Encoding quality.} Even the optimal strategy does not make the MWIS the ground state (rank $= 3$ at freeze, not rank $= 1$). Weighted Rydberg MWIS via local detuning is established \citep{goswami2024,deoliveira2025,yeo2025}; what we measure on Aquila is a rank-3 ceiling under suppression-only local detuning competing with the $C_6/r^6$ tail. The true ground state has one independent-set violation (bitstring \texttt{rgrrgggggggg}, value 214.76 vs.\ MWIS 208.18). The MWIS optimum appears within the bottom-10 eigenstates (rank 3 of 4096, gap $\approx 1.7$ MHz), providing a usable approximate encoding for the hybrid pipeline. This rank-vs-time profile is obtained by a purely classical, exact-diagonalisation sweep (21 evenly spaced $t/T$ values at $N=12$; no sampling, no hardware shots): at each instant the MWIS bitstring's Ising energy is ranked against all 4096 possible bitstrings under that instant's detuning values, as if the system were frozen and measured then. This small-$N$ diagnostic fixes the freeze time \emph{before} any hardware submission---exact diagonalisation is intractable at the deployed $N=50+$ scale, where only final-shot statistics are observable (Section~\ref{sec:rolling}). The encoding window is narrow: enters the bottom-10 at $t/T = 0.85$ and has collapsed by $t/T = 0.95$; the bottom-10 interval itself spans roughly $0.05T \approx 200 ns$ at $T = 4 \microsec$ , and rank 3---not rank 1---is the best value reached anywhere across the full swept range (Figure~\ref{fig:window}). The old pipeline's default parameter set had rank-1 encoding at $t/T = 0.7$, but the freeze occurred at $t/T = 0.9$---$0.15T$ after the encoding collapsed to rank 57.

\section{Results}\label{sec:results}

AHS hardware was engaged for Days 1--15 at 50 nodes, 126--153 edges, and 200 shots per day, plus a follow-up single-day sweep at $N=75/100/125/144$ on Day 1 (Section~\ref{sec:nscaling}). All dispatch validation uses a 100-scenario stochastic LP repair that verifies delivery-contract feasibility and computes realised LP margins. Reported LP margins are per-day standalone values; cumulative rolling state propagation (tank and battery carryover) is reserved for future work.

\subsection{Combinatorial Kernel Performance}\label{sec:classical}

The move-MWIS kernel was evaluated on a chronological-fill reference schedule with 114 of the 144 possible module-hours committed (all 6 modules filled hour-by-hour starting from hour 0, independent of price; the contrasting cheapest-fill policy instead selects the lowest-price module-hour slots first, regardless of hour). LP-dual scoring produced a pruned move set of 50 candidates connected by a conflict graph with 283 edges. Exact MWIS via HiGHS MILP selected an optimal subset attaining a dual-proxy objective of 12.15, compared with 3.63 for a cold-start SA run seeded from an empty move set (Section~\ref{sec:pipeline}). A warm-started SA run, seeded from the HiGHS solution, converged to the same 12.15 optimum at iteration 0: the LP-dual proxy landscape is sharp enough that the exact MWIS solution is also a local optimum under SA's move neighbourhood, and a cold start does not reach it.

A rolling-horizon dispatch configuration---propagating each day's tank and battery terminal state forward as the next day's initial condition---was feasible for a single day, with a positive LP margin of \GBP23{,}622, 100\% dispatch rate, and hydrogen balance residual at machine precision ($\sim 4.5 \times 10^{-13}$).

\subsection{15-Day Rolling AHS Hardware Validation}\label{sec:rolling}

The 15-day rolling evaluation submitted one 50-node, 200-shot AHS task per campaign day to QuEra Aquila via AWS Braket \citep{braket2020}. Identical AHS schedule parameters (the \emph{narrow-zero-cross} recipe, Section~\ref{sec:ahsparams}) and pipeline configuration were used across all 15 days. All 15 scheduled days completed successfully. Extended per-day statistics---conflict-graph size, node-weight range, exact shot counts, discarded shots, pipeline runtime, and the pre-kernel reference LP margin---are reported in Supplementary Section~S3, Table~S3.

\subsubsection{Inner-Kernel: AHS vs.\ Exact MWIS}\label{sec:innerkernel}

Each day's raw AHS bitstring distribution was post-processed through the full 4-phase pipeline: (1) broken-edge conflict resolution, (2) full King-graph independent-set enforcement, (3) negative-weight pruning, and (4) greedy positive-weight-descending maximality augmentation. The post-processed AHS objective value is compared against the exact MWIS optimum (HiGHS MILP) and a greedy heuristic baseline on the same 50-node conflict graph (Table~\ref{tab:inner}).

\begin{table}[htbp]
\centering
\caption{Inner-kernel comparison for Days 1--15.}
\label{tab:inner}
\footnotesize
\setlength{\tabcolsep}{4pt}
\begin{tabular*}{\textwidth}{@{\extracolsep{\fill}}crrrrrr@{}}
\toprule
Day & \thead{AHS-raw\\(\GBP)} & \thead{AHS-post\\(\GBP)} & \thead{AHS-post\\ratio} & \thead{Exact MWIS\\(\GBP)} & \thead{Greedy\\(\GBP)} & \thead{Valid\\rate} \\
\midrule
1 & 5{,}975.70 & 6{,}378.15 & 0.915 & 6{,}972.90 & 6{,}972.90 & 0.720 \\
2 & 3{,}041.75 & 5{,}247.00 & 0.748 & 7{,}017.60 & 6{,}497.20 & 0.595 \\
3 & $-$417.45 & 879.65 & 0.961 & 915.15 & 86.05 & 0.665 \\
4 & $-$3{,}405.00 & 2{,}241.45 & 1.000 & 2{,}241.45 & $-$1{,}006.05 & 0.625 \\
5 & 1{,}297.40 & 5{,}872.00 & 0.985 & 5{,}959.45 & 2{,}896.15 & 0.670 \\
6 & 3{,}346.65 & 5{,}711.85 & 0.989 & 5{,}778.30 & 3{,}847.85 & 0.725 \\
7 & $-$615.15 & 3{,}685.75 & 0.808 & 4{,}561.35 & 1{,}224.75 & 0.545 \\
8 & $-$761.65 & 3{,}940.60 & 0.989 & 3{,}985.40 & 99.75 & 0.705 \\
9 & 1{,}783.65 & 5{,}595.65 & 0.923 & 6{,}061.80 & 3{,}111.60 & 0.625 \\
10 & 314.60 & 2{,}995.20 & 1.000 & 2{,}995.20 & $-$177.75 & 0.625 \\
11 & $-$240.45 & 3{,}011.80 & 0.980 & 3{,}072.00 & $-$95.70 & 0.675 \\
12 & 4{,}570.50 & 5{,}382.40 & 0.894 & 6{,}017.30 & 4{,}180.15 & 0.690 \\
13 & $-$610.85 & 4{,}675.70 & 0.975 & 4{,}795.80 & 2{,}792.70 & 0.645 \\
14 & 1{,}865.45 & 4{,}624.70 & 0.950 & 4{,}869.15 & 2{,}772.10 & 0.655 \\
15 & 3{,}343.30 & 5{,}820.55 & 0.985 & 5{,}911.40 & 4{,}032.35 & 0.685 \\
\bottomrule
\end{tabular*}
\end{table}

\begin{figure}[htbp]
\centering
\includegraphics[width=\linewidth]{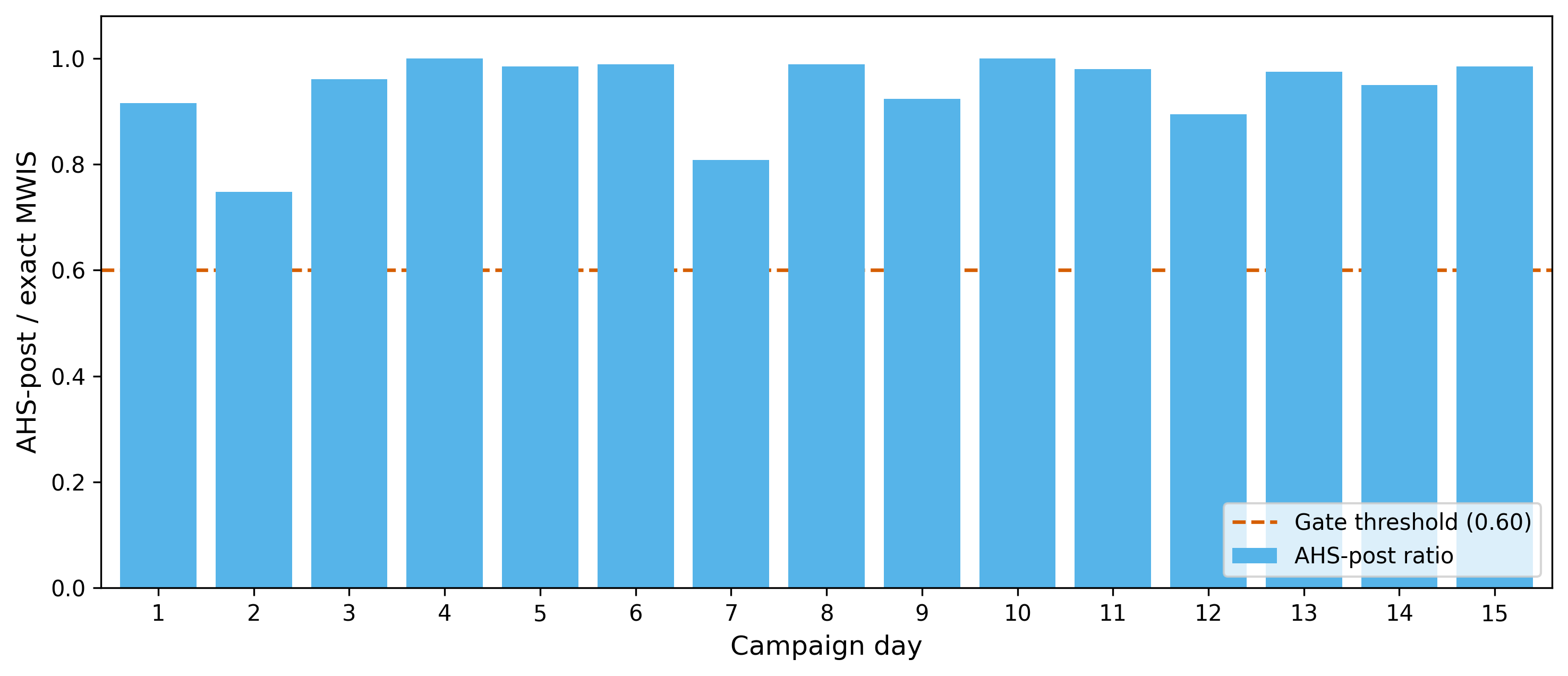}
\caption{AHS-post ratio versus campaign day. Every day exceeds the 0.60 viability threshold.}
\label{fig:rolling}
\end{figure}

\textbf{Aggregate (15 days):} AHS-post ratio mean $= 0.940$, median $= 0.975$, min $= 0.748$ (Day 2), max $= 1.000$. The inner kernel clears its viability threshold (AHS-post ratio $\ge 0.60$) on all 15 days (Figure~\ref{fig:rolling}).

The weakest AHS-post ratio (Day 2, 0.748) does not coincide with the day with the worst raw hardware signal on the full 15-day set: Day 7 has both the lowest valid-shot rate (0.545) and the largest number of discarded shots (91 of 200, 45.5\%), yet still recovers to a 0.808 post-processed ratio.

The raw AHS objective (before post-processing) is strongly influenced by measurement noise and non-conflicting atom-loss artefacts: raw ratios range from $-1.519$ to 0.857 (mean 0.113). Six days (Days 3, 4, 7, 8, 11 and 13) show negative raw objectives driven by surviving negative-weight noise atoms. This occurs because the raw objective is the naive sum of node weights over every atom measured in the Rydberg-excited state, with no independent-set or sign filtering applied: the per-atom detuning $\delta_k(t)$ (Eq.~\eqref{eq:detuning}) biases each atom toward the ground state in proportion to its weight, but this bias is probabilistic rather than a hard constraint---Aquila's local detuning is suppression-only (it can push an atom toward $|g\rangle$ but never actively promote one toward $|r\rangle$) and competes against the uniform global Rabi drive, so finite adiabaticity and measurement noise within the narrow encoding window (Section~\ref{sec:ahsparams}) can still leave a negative-weight atom measured as excited by chance. The 4-phase post-processing pipeline resolves all six days to ratios of 0.808--1.000: its negative-weight-pruning phase deterministically removes every selected atom with $w_v < 0$, a correction that is lossless by construction, since no node with negative weight can ever belong to a true maximum-weight independent set. This confirms that the MWIS-optimal (or near-optimal) bitstring is present in the raw hardware distribution but is masked by non-conflicting negative-weight atoms that the earlier, incomplete post-processing failed to prune.

The greedy heuristic baseline is erratic: it matches (Day 1) or nearly matches (Day 2, 92.6\%) the exact MWIS optimum on the two highest mean-weight days but produces negative objectives on days with predominantly negative node weights (Days 4, 10 and 11), where a naive greedy strategy selects negative-weight nodes early. AHS post-processing outperforms greedy on 13 of 15 days; on the 2 exceptions (Days 1 and 2), greedy's exact-or-near-exact result on those high-mean-weight instances edges out the post-processed AHS selection. Note that phase 4 of post-processing uses the same greedy rule as this standalone baseline, so the comparison isolates the value added by the AHS-selected seed set (phases 1--3) rather than by greedy augmentation itself.

\subsubsection{Total-Kernel: AHS+SA Refinement and LP Dispatch Margin}\label{sec:totalkernel}

The AHS post-processed selection was used as the initial state for a simulated annealing warm-start (MoveSpaceSA, 10{,}000 iterations, initial temperature 100.0, cooling rate 0.999). The SA-refined selection was then dispatched and validated through 100-scenario stochastic LP repair. This total-kernel result is compared against the exact MWIS+LP baseline (the classical upper bound on the same graph) in Table~\ref{tab:total}.

\begin{table}[htbp]
\centering
\caption{Total-kernel comparison: SA-refined LP dispatch margin vs.\ exact MWIS LP dispatch margin for Days 1--15.}
\label{tab:total}
\footnotesize
\setlength{\tabcolsep}{5pt}
\begin{tabular*}{\textwidth}{@{\extracolsep{\fill}}crrrr@{}}
\toprule
Day & \thead{AHS+SA LP\\(\GBP)} & \thead{Exact MWIS LP\\(\GBP)} & \thead{SA\\ratio} & \thead{Margin\\ratio} \\
\midrule
1 & 7{,}193.61 & 7{,}192.92 & 0.927 & 1.000 \\
2 & 8{,}639.34 & 7{,}738.59 & 0.768 & 1.116 \\
3 & 25{,}589.80 & 25{,}589.80 & 1.000 & 1.000 \\
4 & 28{,}359.70 & 28{,}359.70 & 1.000 & 1.000 \\
5 & 7{,}209.82 & 7{,}209.82 & 1.000 & 1.000 \\
6 & 6{,}665.15 & 6{,}665.15 & 1.000 & 1.000 \\
7 & 12{,}820.37 & 12{,}820.36 & 0.808 & 1.000 \\
8 & 12{,}268.14 & 12{,}268.14 & 0.989 & 1.000 \\
9 & 10{,}550.75 & 10{,}250.75 & 0.923 & 1.029 \\
10 & 13{,}587.27 & 13{,}587.27 & 1.000 & 1.000 \\
11 & 12{,}029.78 & 12{,}029.78 & 1.000 & 1.000 \\
12 & 7{,}383.29 & 7{,}083.29 & 0.922 & 1.042 \\
13 & 11{,}909.33 & 11{,}909.33 & 0.975 & 1.000 \\
14 & 13{,}044.39 & 13{,}044.39 & 1.000 & 1.000 \\
15 & 9{,}613.18 & 9{,}613.18 & 1.000 & 1.000 \\
\bottomrule
\end{tabular*}
\end{table}

\begin{figure}[htbp]
\centering
\includegraphics[width=\linewidth]{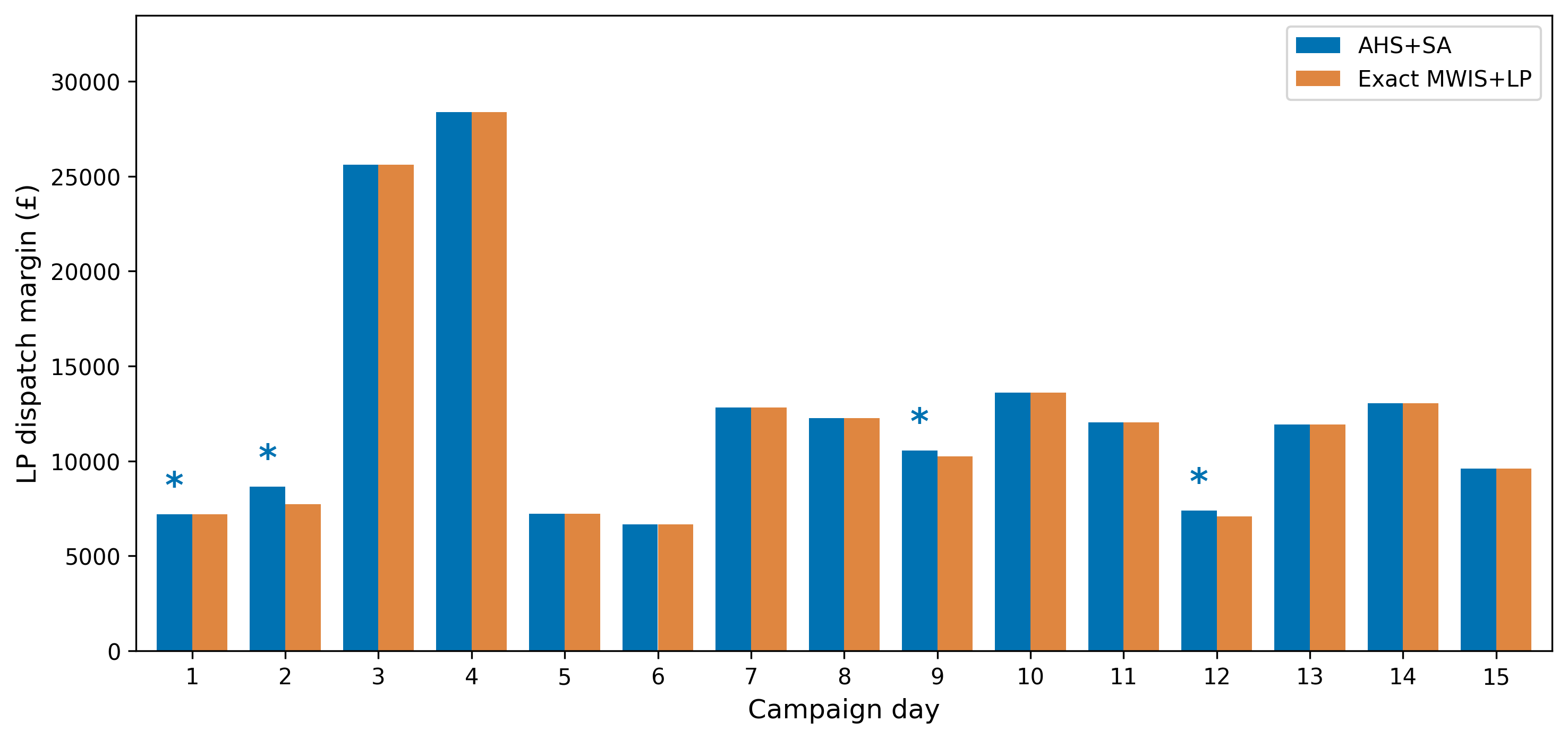}
\caption{Total-kernel LP dispatch margin versus campaign day: AHS+SA-refined vs.\ exact classical MWIS+LP baseline (Table~\ref{tab:total}). Asterisks mark days on which AHS+SA exceeds the exact baseline. The AHS+SA margin matches or exceeds the exact baseline on every one of the 15 rolling days---11 days at exact parity and 4 days with a modest improvement (Days 2, 9 and 12, and the near-parity Day 1 case)---with zero days of underperformance.}
\label{fig:totalkernel}
\end{figure}

\textbf{Aggregate (15 days):} SA-refined objective ratio mean $= 0.954$, median $= 1.000$. SA-refined LP dispatch margin mean $= \GBP12{,}458$, vs.\ exact MWIS LP dispatch margin mean $= \GBP12{,}358$. The SA-refined LP margin matches or exceeds the exact MWIS LP margin on all 15 days: 11 days achieve exact parity (margin ratio $= 1.000$ to the precision shown; Day 1 is marginally higher at full precision, \GBP7{,}193.61 vs.\ \GBP7{,}192.92), and 4 days show a modest improvement (margin ratio 1.029--1.116: Days 2, 9 and 12, and the near-parity Day 1 case). That improvement is not a quantum speedup. It reflects objective mismatch: the MWIS solver maximises the LP-dual proxy, while the dispatch metric is the realised 100-scenario LP margin, a nonlinear function of the selected moves, so SA initialised from the AHS set can leave a proxy local optimum and raise the true margin despite a slightly lower MWIS objective (Figure~\ref{fig:totalkernel}; Section~\ref{sec:finding}).

\subsubsection{AHS Hardware Statistics: Shot Validity}\label{sec:hwstats}

A shot is \emph{valid} when every one of the $N$ requested atoms is present at the pre-evolution occupancy check; shots failing this full-array post-selection are discarded before any bitstring analysis, following standard practice for Aquila data \citep{wurtz2023}. Across all 15 days the mean valid-shot rate was 0.657 (range 0.545--0.725); on average, 68.7 of 200 shots per task (34.3\%, range 55--91) were discarded at the validity check. Under an independent per-atom model $v = p^{N}$, these rates imply an effective per-atom retention of $\hat{p} = v^{1/N} = 0.988$--$0.994$ (mean 0.992; Supplementary Section~S4, Table~S4), of the same order as, though modestly below, the published exponential sorting-fidelity bound of $\approx 0.995^{N}$ for this device class \citep{kaufman2026}. Here $\hat{p}$ aggregates loading, rearrangement, vacuum loss, and imaging into one effective rate; it is not a channel decomposition. Every submitted layout passed the field-of-view check, consistent with fitting within Aquila's $75\times76$~\micron\ FOV under the 2-row folding embedding (Section~\ref{sec:embedding}). Fourteen of 15 days have zero inter-row broken edges, because the span-minimising split point (Eq.~\eqref{eq:split}) lands on a natural gap in that day's occupied-hour distribution. The remaining day (Day 4) has no such gap and retains 16 broken edges---still resolved to the exact MWIS optimum by the 4-phase post-processing pipeline (Table~\ref{tab:inner}, AHS-post ratio 1.000). The shot-discard rate correlates weakly with the post-processed ratio ($r = -0.56$): higher discard reduces the effective sample size but does not remove the MWIS bitstring from the valid-shot distribution when the schedule is well tuned.

\subsubsection{Parameter Tuning: Sim Sweep Results}\label{sec:paramtuning}

The 11-strategy AHS schedule parameter sweep at $N=12$ (exact diagonalisation, $2^{12} = 4{,}096$ states) was reported in Section~\ref{sec:ahsparams} and is tabulated and plotted in Supplementary Section~S2 (Table~S2, Figure~S1). A follow-up 3-candidate sim sweep tested whether the encoding could be improved beyond the baseline \emph{narrow-zero-cross} recipe. All three candidates (freeze at $t/T = 0.85$ instead of 0.90, wider symmetric sweep $\pm 40$~MHz, and Rabi frequency variation at 7 or 9~MHz) produced MWIS rank $\ge 3$ at $N=12$, with no improvement over the baseline (rank $= 3$, gap $\approx 1.7$ MHz). Hardware validation of alternative parameters was therefore skipped. The \emph{narrow-zero-cross} recipe remains the sole empirically validated AHS schedule for this pipeline.

\subsubsection{N-Scaling: Extending Beyond N=50}\label{sec:nscaling}

The 15-day rolling campaign fixed $N=50$ (the top-$k$ pruning cutoff, Section~\ref{sec:pipeline}) across all days. A separate single-day sweep on Day 1 (the same reference instance and \emph{narrow-zero-cross} schedule as the rolling campaign) tests whether the 2-row folding embedding continues to hold as $N$ grows. With 6 modules and a 24-hour horizon, the move vocabulary's one-candidate-per-cell scheme has a hard ceiling of $N = 6 \times 24 = 144$ distinct (module, hour) grid cells; $N=150$ and beyond are not reachable without changing the reference schedule's module or hour shape, which is out of scope for this sweep. Four new values were submitted to real Aquila hardware---$N=75$, 100, 125, 144---each at 200 shots, alongside the existing $N=50$ data point from the rolling campaign (Table~\ref{tab:inner}, Day 1 row). Results are given in Table~\ref{tab:nscaling}.

\begin{table}[htbp]
\centering
\caption{$N$-scaling sweep on Day 1, 200 shots per $N$. Each instance uses $N$ atoms (one atom per retained move).}
\label{tab:nscaling}
\footnotesize
\setlength{\tabcolsep}{4pt}
\begin{tabular*}{\textwidth}{@{\extracolsep{\fill}}rrrrrrrr@{}}
\toprule
$N$ & \thead{Valid\\rate} & \thead{AHS-post\\(\GBP)} & \thead{Exact MWIS\\(\GBP)} & \thead{Greedy\\(\GBP)} & \thead{AHS-post\\ratio} & \thead{Broken\\edges} & $x_{\max}$ (\micron) \\
\midrule
50  & 0.720 & 6{,}378.15  & 6{,}972.90  & 6{,}972.90  & 0.915 & 0  & 35.0 \\
75  & 0.455 & 8{,}784.85  & 9{,}247.80  & 7{,}925.40  & 0.950 & 0  & 40.0 \\
100 & 0.095 & 9{,}504.95  & 10{,}608.75 & 9{,}274.95  & 0.896 & 16 & 50.0 \\
125 & 0.125 & 12{,}124.25 & 12{,}766.35 & 11{,}240.10 & 0.950 & 16 & 55.0 \\
144 & 0.245 & 12{,}020.45 & 13{,}636.35 & 11{,}240.10 & 0.881 & 16 & 55.0 \\
\bottomrule
\end{tabular*}
\end{table}

\begin{figure}[htbp]
\centering
\includegraphics[width=\linewidth]{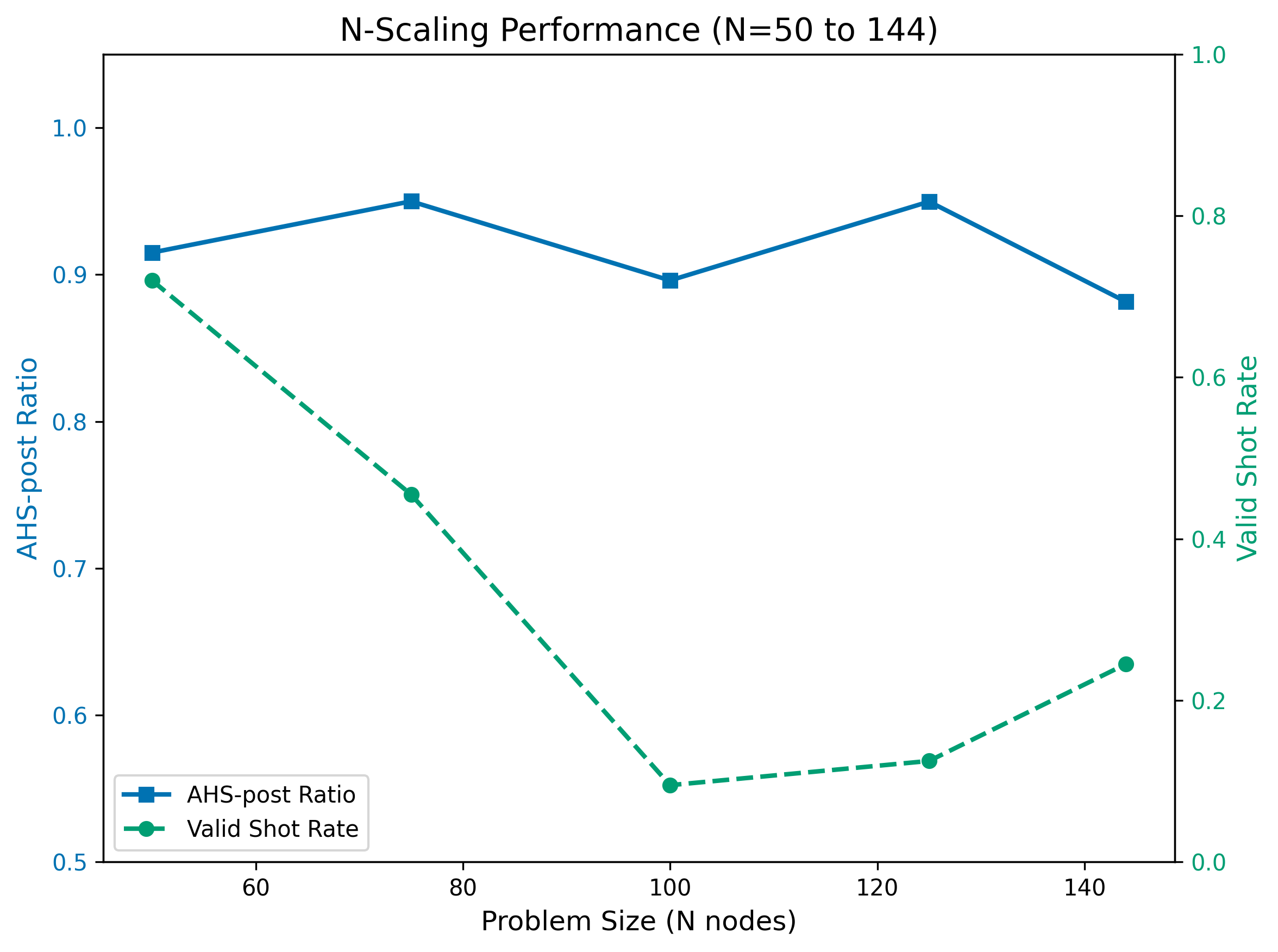}
\caption{$N$-scaling from $N=50$ to $N=144$. Valid-shot rate collapses with atom count; FOV occupancy and AHS-post ratio do not.}
\label{fig:nscaling}
\end{figure}

Across a near-threefold increase in problem size ($N=50$ to $N=144$), FOV geometry is not the limit: $x_{\max}$ grows from 35 to 55~\micron\ (75~\micron\ hardware cap) and $y_{\max}$ is flat at 64~\micron\ (76~\micron\ cap). The AHS-post ratio shows no systematic degradation with $N$, ranging 0.881--0.950 (mean 0.918). The valid-shot rate does degrade: it falls from 0.720 at $N=50$ to 0.095 at $N=100$ (19 of 200 shots), then recovers only to 0.245 at $N=144$. Per-shot validity, not atom-count headroom (256) or FOV, is the binding operational limit for this AHS optimisation. Under the $v = p^{N}$ model, the sweep's implied per-atom retention is $\hat{p} = 0.977$--$0.990$ (Supplementary Section~S3, Table~S3); the non-monotonic recovery at $N=144$ reflects task-level variation in $\hat{p}$, which the exponent amplifies---at $N=100$, $\hat{p} = 0.977$ yields $\approx 10\%$ of shots surviving where $\hat{p} = 0.990$ (the $N=144$ task's value) would yield 37\%. Even those 19 valid shots at $N=100$ recover a 0.896 post-processed ratio. Sixteen broken edges appear starting at $N=100$ and remain flat through $N=144$, because the span-minimising split continues to land near the same natural gap. AHS-post exceeds the greedy baseline at every new $N$ tested; the sole exception is the $N=50$ reference, already established in Table~\ref{tab:inner} as one of the two days where greedy matches or beats AHS-post on this day's weight distribution. Figure~\ref{fig:nscaling} summarises the trend.

\section{Discussion}\label{sec:discussion}

\subsection{End-to-end stochastic UC on Aquila}\label{sec:contribution}

The closed loop is UC$\to$move-MWIS$\to$Aquila AHS$\to$100-scenario LP repair, demonstrated on a green-hydrogen electrolyser park. Prior quantum UC is annealer-side QUBO plus LP \citep{barrass2025,christeson2025}. Solver-level studies benchmark those annealers against classical MIP \citep{quinton2025,ribes2026}. Prior Aquila power-system work is MaxCut on graphs of size $\le 12$, with no commitment layer and no dispatch repair \citep{bauer2026}. To our knowledge this is the first neutral-atom UC encoding with an application-level repair (Table~\ref{tab:positioning}).

The King-graph used on hardware is a conservative compatibility proxy, not the UC constraint set. Feasibility is enforced downstream: every reported day returns a 100-scenario LP margin, so the geometric approximation is absorbed by repair rather than asserted as a physical scheduling rule (Section~\ref{sec:embedding}).

\subsection{Fifteen-day campaign and proxy--realised mismatch}\label{sec:finding}

Over 15 consecutive cloud days the hybrid AHS+SA pipeline matches or exceeds exact MWIS+LP on the dispatch metric: mean \GBP12{,}458 vs.\ \GBP12{,}358, with four days improved by 2.9--11.6\%. Inner-kernel AHS-post retains a mean 94.0\% of the exact MWIS objective (range 74.8--100\%) after 4-phase post-processing. The campaign is the evaluation unit: one 50-node, 200-shot task per day, identical schedule parameters, all 15 days completed.

Where AHS+SA beats exact MWIS on LP margin, the cause is objective mismatch, not a quantum speedup. Exact MWIS maximises the LP-dual proxy; the figure of merit is the realised stochastic LP. A noisy AHS sample plus classical SA can therefore improve the true objective while scoring lower on the proxy. HiGHS solves the same 50-node MWIS in under a second. The operational test is whether a NISQ sampler degrades the repaired dispatch margin relative to that baseline. For this case study, over 15 days, it does not.

\subsection{Per-shot validity as the binding AHS scale limit}\label{sec:shots}

FOV folding works: $x_{\max}$ stays at 35--55~\micron\ against a 75~\micron\ cap out to $N=144$, and 14 of 15 campaign days have zero broken inter-row edges. Encoding quality is likewise stable (AHS-post 0.881--0.950). What does not scale is the valid-shot rate, from 0.720 at $N=50$ to 0.095 at $N=100$; at campaign scale, 34.3\% of shots per task (55--91 of 200) were discarded at the full-array validity check. The collapse is expected in kind: Aquila fills static traps with $\sim$60\% probability and post-selects on fully loaded arrays \citep{wurtz2023}, and device diagnostics model the resulting sorting fidelity as $\approx 0.995^{N}$ \citep{kaufman2026}, so exponential post-selection in atom number is built into the fill-and-sort architecture. What the campaign adds is the deployment-scale quantification: across all 20 hardware tasks the implied per-atom retention is $\hat{p} = v^{1/N} = 0.977$--$0.994$ (Supplementary Section~S3, Table~S3), slightly below the published bound, so the collapse is the arithmetic of full-array post-selection rather than a device shortfall---even at $p = 0.995$, only $\approx 49\%$ of shots survive at $N=144$ and $\approx 28\%$ at the 256-atom maximum. These rates are specific to the present layouts and to full-array post-selection; a channel-resolved loss analysis (loading, vacuum lifetime, imaging, rearrangement, reservoir depletion) would be needed before reading the 0.095 figure as a device-wide constant.

Few valid shots still suffice for the inner kernel once the schedule is fixed: 19 shots at $N=100$ recover an AHS-post ratio of 0.896. Shot-scaling studies on Rydberg arrays report near-constant shot requirements at relaxed approximation targets because sampled distributions concentrate near good solutions \citep{jung2026}; the mechanism here is different: 4-phase post-processing repairs individual bitstrings, so a handful of valid shots suffices provided one lies in the repairable neighbourhood of a near-optimum.

\subsection{Kernel: scoring, folding, and a concrete IR}\label{sec:embeddingdiscussion}

Three compilation choices underpin the end-to-end demonstration. LP-dual weights come from one 100-scenario stochastic LP; candidates are then scored in closed form, with no per-move re-solve. Two-row folding embeds the 24-hour lattice with no gadget ancillae, against $O(N^2)$ universal constructions \citep{nguyen2023}, by splitting time at the span-minimising gap. The emitted conflict graph is the IR that was solved both by HiGHS and by Aquila without rewriting the kernel; annealing and coherent Ising machines can take the same graph \citep{lucas2014,wurtz2024hybrid}.

On Aquila, local detuning is suppression-only. Combined with the $C_6/r^6$ tail, that yields a rank-3---not rank-1---MWIS encoding at $N=12$ \citep{goswami2024,deoliveira2025,yeo2025}. Within the \emph{narrow-zero-cross} recipe the encoding is invariant in $T$ over 2--6~\microsec, because the freeze sits in a $\sim$200~ns window in $t/T$. After post-processing, mean AHS-post 0.940 is enough to drive SA and LP repair. The true Ising ground state at freeze has an independent-set violation and is not a feasible MWIS.

\subsection{Outlook}\label{sec:outlook}

The $N=50$--144 instances are classically easy King graphs \citep{andrist2023,schuetz2025,cazals2026}. Unweighted kernelization and later weighted MWIS codes \citep{lamm2017,lamm2019mwis,grossmann2024} would not change the exact-MILP baseline at this scale. A quantum role, if any, begins only on a larger reference instance (more modules or a longer horizon), where shot validity will become limiting before the FOV does.

The hydrogen model is an electrical conversion factor, not a multiphysics plant. Thermal, impurity, and part-load constraints \citep{qiu2023}, chance-constrained duals \citep{ribes2026}, and carbon or time-matching metrics \citep{giovanniello2024} still compile to the same weighted MWIS graph. Multi-cell moves remain classical; restoring them on hardware would need gadget or parity encodings \citep{nguyen2023,lanthaler2023}. Quantum wires are the lower-overhead route when only a few non-local edges appear \citep{kim2022wires,deoliveira2026wires}. The next hardware data point is the same graph on a quantum annealer \citep{quinton2025,ribes2026} or a coherent Ising machine \citep{takesue2025}.

\section{Conclusion}\label{sec:conclusion}

A move-MWIS kernel compiles stochastic unit commitment into a weighted conflict graph, scored from one LP, solved classically or on analog hardware, and repaired by a 100-scenario dispatch LP. The industrial case is a six-module green-hydrogen electrolyser park.

The pipeline is closed: UC$\to$MWIS$\to$Aquila$\to$LP repair. Over 15 cloud days at $N=50$, AHS+SA dispatch margins match or exceed exact MWIS+LP (mean \GBP12{,}458 vs.\ \GBP12{,}358; four days $+2.9$--$11.6\%$), because the true LP objective is not identical to the dual-proxy MWIS objective. Two-row folding keeps the layout inside Aquila's FOV with zero gadget ancillae (14 of 15 days have no broken edges). Encoding quality holds to $N=144$ (AHS-post 0.881--0.950); valid-shot rate does not (0.720 to 0.095 at $N=100$). Per-shot validity, not field of view, is the binding AHS scale limit.

The instances are classically easy. The campaign measures pipeline fidelity on a repaired industrial dispatch loop, not computational speedup.

The kernel, figure scripts, and hardware-result files are released at \url{https://github.com/Cambridge-EQ/mwis-uc-kernel}. The same weighted conflict graph is the input format for quantum annealing and coherent Ising machines; the next hardware data point is the same graph on one of those backends at a scale where full-array post-selection remains affordable.

\section{CRediT authorship statement }\label{sec:authorship}

\textbf{Jiying Chen}: Conceptualization, Methodology, Software, Investigation, Validation, Writing -- original draft, Writing -- review \& editing. \textbf{Min Lin}: Methodology, Software, Investigation, Validation, Writing -- original draft. \textbf{Jingwei Wen}: Visualization, Writing -- review \& editing. \textbf{Zhihong Zhang}: Visualization, Writing -- review \& editing. \textbf{Chuixiong Wu}: Conceptualization, Methodology, Investigation, Visualization, Writing -- original draft, Writing -- review \& editing. 

All authors reviewed and approved the final manuscript and confirmed the accuracy of their stated contributions. All quantum-hardware experiments were conducted exclusively by Jiying Chen and Min Lin in Cambridge, United Kingdom, using personal AWS accounts to access quantum processors through the commercially available Amazon Braket service.

\bibliographystyle{unsrtnat}
\bibliography{bibliography}

\end{document}